\documentclass[12pt]{iopart}

\usepackage{iopams}

\RequirePackage{hyperref}

\newcommand{\pd}{\partial}

\begin{document}

\title{Effective Potential of Scalar-Tensor Gravity}

\author{\href{https://orcid.org/0000-0001-9326-6905}{Andrej Arbuzov} }
 
\address{Bogoliubov Laboratory for Theoretical Physics, JINR, Dubna 141980, Russia \\
Dubna State University, Universitetskaya str. 19, Dubna 141982, Russia}

\ead{\href{mailto:arbuzov@theor.jinr.ru}{arbuzov@theor.jinr.ru}}

\author{\href{https://orcid.org/0000-0001-7099-0861}{Boris Latosh} }
\address{Bogoliubov Laboratory for Theoretical Physics, JINR, Dubna 141980, Russia \\
Dubna State University, Universitetskaya str. 19, Dubna 141982, Russia}
\ead{\href{mailto:latosh@theor.jinr.ru}{latosh@theor.jinr.ru}}

\vspace{10pt}
\begin{indented}
\item[]%June 2020
\end{indented}

\begin{abstract}
  Effective potential of a scalar field induced by weak gravity is studied. The set of operators providing the leading contribution and preserving the second order of field equations is found. It is shown that only a mass term and a specific Brans-Dicke-like interaction are relevant within such a setup. An explicit form of the potential is found. The model has room for a natural inflationary scenario similar to the well-known case of the Starobinsky inflation. Possible implications for the Standard Model are highlighted.
\end{abstract}

\vspace{2pc}

\submitto{\CQG}

%\maketitle

\section{Introduction}

Scalar-tensor models of gravity occupy a special place in the modified gravity landscape. They extend the gravitational sector with an additional scalar degree of freedom providing, perhaps, the simplest alternative to general relativity. Horndeski theories are the most general class of models admitting second order field equations~\cite{Horndeski:1974wa,Kobayashi:2011nu}. Despite their apparent simplicity, scalar-tensor theories have applications in many areas such as cosmology, inflation, black hole physics, etc.~\cite{Ishak:2018his,Berti:2015itd,Clifton:2011jh,Tretyakova:2017fbu}.

The structure of a scalar field potential is crucial for a theory, as it provides
room for some important physical phenomena. Its role is well-illustrated with two following examples. Firstly, within inflation theory the form of a potential defines whether 
a model admits the slow-roll inflation~\cite{Linde:1983gd,Senatore:2016aui,Gorbunov:2011zzc}.
Secondly, some scalar-tensor models experience so-called Chameleon screening~~\cite{Khoury:2003rn,Khoury:2003aq,Brax:2008hh,Burrage:2017qrf}. The screening requires an existence of of a specific non-minimal coupling between the scalar field and matter together with a potential of a special form.
%models which admit both a non-minimal coupling between the scalar field and matter and have a potential of a certain form experience the so-called Chameleon screening~\cite{Khoury:2003rn,Khoury:2003aq,Brax:2008hh,Burrage:2017qrf}.
Due to the non-minimal coupling a scalar field potential receives an additional contribution 
proportional to the local matter density, develops a large effective mass, and ceases to propagate.

A scalar field potential is altered at the quantum level as the theory develops an effective potential 
that can radically change its features~\cite{Coleman:1973jx,Buchbinder:1992rb}. In the simplest case 
an effective potential develops a new independent energy scale in the infrared sector and breaks 
the conformal symmetry of a model~\cite{Coleman:1973jx}. Geometry of a curved spacetime also 
affects the effective potential~\cite{Buchbinder:1992rb}. This makes it essential to study 
the influence of quantum gravitational effects on a scalar field potential. 

The effective field theory approach provides a framework capable to account for quantum gravitational 
effects and to avoid issues related to the non-renormalizable nature of quantum 
gravity~\cite{Burgess:2003jk,Donoghue:1994dn}. This approach was applied for gravity before 
and was found to be fruitful~\cite{Bjerrum-Bohr:2014zsa,Heisenberg:2020cyi,Calmet:2018qwg,Latosh:2018xai,Arbuzov:2017nhg}.

In this paper we address the influence of quantum gravitational effects on a scalar field potential. 
The paper is organized as follows. In Section~\ref{section_the_potentials} we highlight the leading 
corrections relevant for an effective potential. We show that a massless scalar field minimally coupled 
to gravity does not develop an effective potential. The simplest non-minimal coupling affecting 
an effective potential belongs to Horndeski models and resembles the coupling from the Brans-Dicke 
theory. In Section~\ref{section_implementations} we discuss possible implementations of these results. Namely, we point out that the effective potential may become relevant for inflation theory, 
as it can fit the slow-roll conditions. We also argue that the Chameleon screening can hardly be 
relevant within such a setup as quantum effects can be safely neglected in a matter dense environment. 
Finally, we point to some possible relations with a spontaneous conformal symmetry breaking and the 
cosmological constant problem. We conclude in Section~\ref{section_conclusion}.

\section{Scalar field effective potential}\label{section_the_potentials}

The most general class of scalar-tensor models admitting second-order field equations is given by 
the Horndeski Lagrangians~\cite{Horndeski:1974wa,Kobayashi:2011nu}:
\begin{eqnarray}\label{Horndeski_Lagrangians}
  \eqalign{\mathcal{L}_2 = G_2 (\phi,X) , \cr
  \mathcal{L}_3 = G_3 (\phi,X)\,\square\phi, \cr
  \mathcal{L}_4 = G_4 (\phi,X) R + G_{\rm 4,X} \left[ (\square\phi)^2-(\nabla_{\mu\nu}\phi)^2 \right], \cr
  \mathcal{L}_5 = G_5 (\phi,X)\,G^{\mu\nu}\nabla_{\mu\nu}\phi-\frac16\,G_{\rm 5,X} \left[ (\square\phi)^3 -3 (\square\phi)(\nabla_{\mu\nu}\phi)^2 + 2 (\nabla_{\mu\nu}\phi)^3 \right].}
\end{eqnarray}
Here $X=\frac12\,(\pd\phi)^2$ is the standard kinetic term of a scalar field; $G_{\rm i}$ are arbitrary smooth functions; $G_{\rm i,X}$ are the corresponding derivatives with respect to $X$; $R$ is the scalar curvature,
and $G_{\mu\nu}$ is the Einstein tensor. Horndeski Lagrangians~\eref{Horndeski_Lagrangians} define 
the structure of a theory. Terms $\mathcal{L}_4$ and $\mathcal{L}_5$ constrain the spectrum of 
a suitable non-minimal interaction with gravity while $\mathcal{L}_2$ and $\mathcal{L}_3$ specify 
the structure of a scalar field self-interaction.

Since Horndeski models admit second-order field equations, they are free from instabilities associated 
with higher derivatives. Therefore they are highlighted by the stability reasoning. 
It should be noted that there is a generalization of Horndeski models known as {\em beyond Horndeski} 
which also admits second order field equations. These models obtained from Horndeski Lagrangians 
via disformal transformations~\cite{Zumalacarregui:2013pma,Bekenstein:1992pj,Kobayashi:2019hrl}. 
These transformations map beyond Horndeski models with the minimal coupling to matter on Horndeski 
models with non-minimal couplings. In other words, beyond Horndeski models extend interactions with the regular matter. The main focus of this paper is on the gravitational 
influence on an effective potential, henceforward we will discuss only Horndeski models. 

We pursuit the goal to study the most universal effects taking place due to quantum gravitational
contributions. The effective potential, alongside other quantum effects, strongly depends on 
the structure of a given model. To account only for the universal effects, we focus on models without 
scalar field self-interactions and make only a brief comment on their account. For the sake of 
simplicity, we also separately discuss the cases with minimal and non-minimal couplings to gravity.

Application of effective field theory formalism for gravity models is widely covered in literature \cite{Buchbinder:1992rb,Burgess:2003jk,Donoghue:1994dn,Barvinsky:1985an} (see also \cite{Rivat:2019xrq} for a more general discussion of effective theories in the context of renormalization). An effective theory for gravity is defined in a narrow energy interval $0 \leq E \leq \mu$. Here $\mu$ is called the renormalization scale and it lies far below the Planck mass $m_P$. A theory is given by its microscopic action $\mathcal{A}$ which is defined at the renormalization scale $\mu$. The theory is extended in the low-energy area by the standard means of loop corrections. In such an approach the problem of divergent contribution is soften. First of all, an effective theory is defined in a narrow energy band and cannot be used for arbitrary large energies. Secondly, all divergent contributions can be normalized, the corresponding finite valued of parameters are taken either from the low-energy ($E\sim 0$) or from the high-energy region ($E\sim \mu$). To put it otherwise, the renormalization problem is soften by the price of a finite applicability domain.

For a given gravity model an effective theory is constructed as follows. First of all, the microscopic action $\mathcal{A}$ is defined at the normalization scale. It is naturally to expect that a model is gauge-invariant with respect to coordinate transformations. Therefore we assume that the gauge is fixed at this level. Secondly, a suitable background metric $\overline{g}_{\mu\nu}$ that minimize $\mathcal{A}$ is chosen. In turn, gravity is described with small metric perturbations $h_{\mu\nu}$ propagating about the background metric $\overline{g}_{\mu\nu}$, so the full spacetime metric $g_{\mu\nu}$ reads:
\begin{equation}
  g_{\mu\nu} = \overline{g}_{\mu\nu} + h_{\mu\nu} .
\end{equation}
Therefore a given gravity model (that also may contain matter degrees of freedom) is described with the following generating functional \cite{Burgess:2003jk,Donoghue:1994dn}:
\begin{equation}
  \fl Z = \int\mathcal{D}[h]\mathcal{D}[\psi] \exp\left[i \mathcal{A}[\overline{g},\Psi] + i\frac{\delta \mathcal{A} [\overline{g}, \Psi]}{\delta \overline{g}_{\mu\nu}} h_{\mu\nu} + i \frac{\delta^2\mathcal{A} \,[\overline{g} , \Psi]}{\delta \overline{g}_{\mu\nu}\, \overline{g}_{\alpha\beta} }\, h_{\mu\nu}\, h_{\alpha\beta} + \mathcal{O}(h^3) \right] .
\end{equation}
Here $\Psi$ notes matter degrees of freedom. The first term on the right-hand side corresponds to the matter sector of a theory. The second term describes interactions between matter and gravity. The third terms describes propagation of free gravitational field perturbations. Finally, $\mathcal{O}(h^3)$-terms describe self-interactions of gravitational perturbations and higher-order interaction with matter.

The main interest of this paper is corrections generated by quantum gravitational effects. Because of this, we will consider quantum perturbations of gravitational field (which we will call gravitons for the sake of simplicity). The corresponding effective theories are defined below the Planck scale, so $\mathcal{O}(h^3)$ are negligible and will be omitted in this paper. In other words, one may say, that within such an approach only gravity is quantized while matter fields remain classical.

This approach allows one to use well-developed tools of effective field theory \cite{Donoghue:1994dn,Barvinsky:1985an,Buchbinder:1992rb}. Moreover, it allows one to avoid complex calculations occurring in mixed metric-scalar theories \cite{Barvinsky:1993zg,Shapiro:1995yc,Steinwachs:2011zs,Kamenshchik:2014waa}.

We use the simplest method based on direct perturbative calculations to find a one-loop effective potential~\cite{Coleman:1973jx,Buchbinder:1992rb}. Firstly, one recovers one-loop connected 
irreducible $n$-point Green functions $\mathcal{G}_n$.
Secondly, an $n$-point vertex function $\Gamma_n$ is obtained form $\mathcal{G}_n$ via an amputation of external lines. In $\Gamma_n$ 
all external momenta are set to zero to account only for potential interactions (i.e. interactions that do not depend on particles momenta). In such a setup, 
the effective potential is given in terms of vertex functions $\Gamma_n$ as follows~\cite{Buchbinder:1992rb}:
\begin{equation}
    V_{\rm eff}(\phi) = i \, \sum\limits_{n=1}^\infty\,\frac{1}{n}\,\Gamma_n\,\phi^n.
\end{equation}

Let us address the simplest case of a scalar field without self-interactions coupled to gravity 
in the minimal way:
\begin{equation}\label{the_minimal_model}
    \mathcal{A}_0=\int d^4 x \sqrt{-g}\,\left[-\frac{2}{\kappa^2}\,R+\frac12\,g^{\mu\nu}\, \pd_\mu\phi\, \pd_\nu\phi -\frac{m^2}{2}\,\phi^2 \right],
\end{equation}
with $\kappa^2=32\pi G_N$ begin related to the Newton gravitational constant $G_N$.
The expansion of action $\mathcal{A}_0$ in a series with respect to small metric perturbations $h_{\mu\nu}$ propagating over the flat spacetime $\eta_{\mu\nu}$ contains an infinite number of terms.
In the low-energy area, where an effective theory takes place, it is essential to study only the leading order effects, as they themselves are suppressed by the Planck mass.
Higher orders of perturbation theory are suppressed 
even stronger by higher powers of the Planck mass and can be omitted for the time being. 
It is worth noting that the same setup should be applied for strong gravity phenomena, 
like black holes or the early Universe, with caution, as higher-dimension operators may no longer be neglected and the effective theory setup may lost its relevance. Finally, we use the harmonic gauge throughout the paper with the following gauge-fixing term:
\begin{equation}
  \mathcal{A}_{\rm gf} = \int d^4 x \left[ \pd_\mu h^{\mu\nu} - \frac12\, \pd^\mu h_\sigma^{~\sigma} \right]^2.
\end{equation}
Feynman rules for models addressed in this paper are discussed in details in\cite{Latosh:2020zho}.

In the massless case $m=0$, all vertex functions $\Gamma_n$ vanish. This is due to the fact that 
the scalar field energy-momentum tensor is bilinear in derivatives. Because of this, if at least 
one external momentum in $\Gamma_n$ vanishes, so does the vertex.
This leads to the first important result: {\em a massless scalar field 
itself does not generate an effective potential due to quantum gravitational corrections at 
the one-loop level}.

In the case of a non-vanishing mass $m\not =0$, an effective potential is generated.
In can be seen that vertex functions with an odd number of fields vanish due to the structure of the interaction with gravity. The gravity is coupled to an energy-momentum tensor which is bilinear both in derivatives and fields. Because of this the gravitational interaction cannot change the number of matter states. On the other hand, an arbitrary $n$-point diagram can be viewed as an amplitude of a scattering process. As it was just noted, the number of scalar particles is presented in such processes. Therefore all diagrams with an odd number of fields vanish. A vertex function with $2n$ fields can be evaluated explicitly with the standard technique \cite{Buchbinder:1992rb}:
\begin{equation}
  \Gamma_{\rm 2n} = \mu^{4-d}\int\frac{d^dk}{(2\pi)^d} \,\left(\frac{\kappa}{2}\right)^{2n}\, \left(\frac{m^2}{k^2} \frac{m^2}{k^2-m^2}\right)^n\, \left(1-\frac{d}{2}\right)^n\,d^n .
\end{equation}
Dimensional regularization was used here to evaluate this expression with $\mu$ being the normalization energy scale. The corresponding effective potential reads
\begin{equation}\label{the_first_effective_potential}
  \fl\eqalign{ (V_0)_{\rm eff} &= i\, \mu^{4-d} \int\frac{d^d k}{(2\pi)^d} \left[-\frac12\,\ln\left(1-\frac{\kappa^2}{4}\,\frac{m^2}{k^2}\,\frac{m^2}{k^2-m^2}\, d\,\left(1-\frac{d}{2}\right) \,\phi^2 \right)\right] \cr
    & = \frac{m^4 \kappa^2 \phi^2}{16\pi^2}\,\Bigg[ -\frac{1}{d-4}-\frac{\gamma}{2}+\frac12\,\ln(8\pi) -\frac12\ln\frac{m^2}{\mu^2}-\frac{\ln(2)}{4\kappa^2\phi^2} \cr
      & ~~~ +\left(\frac{1}{8\kappa^2\phi^2}\left(1-\sqrt{1-4\kappa^2\phi^2}\right)-\frac14\,\right) \ln\left[1-\sqrt{1-4\kappa^2\phi^2}\right] \cr
      &~~~ +\left(\frac{1}{8\kappa^2\phi^2}\left(1+\sqrt{1-4\kappa^2\phi^2}\right)-\frac14\,\right) \ln\left[1+\sqrt{1-4\kappa^2\phi^2}\right] \Bigg] + \mathcal{O}(d-4) . }
\end{equation}
The expression is valid only for small values of $\phi$ for which $4 \kappa^2\phi^2 \ll 1$. 
The effective potential should be renormalized on the observable mass $m_{\rm obs}$ measured at the low energy regime % normalization energy scale $\mu$
via the following counterterm:
\begin{equation}
  \delta\mathcal{L} =\frac{m_{\rm obs}^2}{2}\,\phi^2 -  \frac{m^4 \kappa^2 \phi^2}{16\pi^2} \left[-\frac{1}{d-4} -\frac{\gamma}{2}-\frac14+\frac12\,\ln(4\pi) - \ln\frac{m}{\mu}\right].
\end{equation}
It should be highlighted that terms containing $\log$-functions provide a finite local contribution 
to the mass term, so their contribution is accounted in the counterterm. The renormalized effective potential is given by the following expression:
\begin{eqnarray}\label{the_first_effective_potential_ren}
  \fl\eqalign{ (V_0)_{\rm eff,ren}=& - m^4 \,\frac{\ln(2)}{64 \pi^2} + \frac{m_{\rm obs}^2}{2}\,\left[1+\frac{m^2}{m_{\rm obs}^2}\,\frac{m^2\kappa^2}{32\pi^2}\, (1+\ln(4))\right] \,\phi^2  \cr
    &+\frac{m^4}{128\pi^2}\Bigg[\left(1-\sqrt{1-4\kappa^2\phi^2}  -2 \kappa^2 \phi^2\,\right)\ln\left[1-\sqrt{1-4\kappa^2\phi^2}\right] \cr
      &+\left(1+\sqrt{1-4\kappa^2\phi^2}-2 \kappa^2 \phi^2\,\right)\ln\left[1+\sqrt{1-4\kappa^2\phi^2}\right] \Bigg]. }
\end{eqnarray}
{\em This renormalized effective potential~\eref{the_first_effective_potential_ren} constitutes 
  the second result of this paper}. It is generated in the minimal model, so it appears universally in any other theory which contains a scalar of a non-vanishing mass.
The potential contains two mass parameters. The first one $m$ we call the tree-level mass, as it is generated by the microscopic action. This mass parameter is measured at the normalization scale $\mu$ where the microscopic action is given. The second mass parameter $m_{\rm obs}$ we call the observed mass. It is measured in the low energy regime, in $\phi \sim 0$ area. It corresponds to the mass of small perturbations existing far below the normalization scale. 

Properties of this potential should be discussed. Firstly, both~\eref{the_first_effective_potential} 
and~\eref{the_first_effective_potential_ren} are regular about $\phi \sim 0$ and vanish at $\phi=0$. 
This goes in line with the well-known case~\cite{Coleman:1973jx}, as the scalar field mass $m$ serves 
as a natural regulator of the infrared sector. Secondly, before a normalization the 
potential~\eref{the_first_effective_potential} vanishes in the $m\to 0$ limit in full agreement 
with {\em the first result} found above. Finally, potential~\eref{the_first_effective_potential_ren} 
is sensitive to the hierarchy of mass scales. This can be clearly seen from the following expression 
given in term of the tree-level mass $m$, the observed mass $m_{\rm obs}$, and the Planck mass $m_{\rm P}$:
\begin{equation}\label{the_first_effective_potential_ren_series}
  \fl (V_0)_{\rm eff,ren}=\frac{m_{\rm obs}^2}{2}\,\phi^2 +8 \frac{m^4}{m_{\rm P}^4}
  \left(1-\ln(4)+2\ln\left[64 \pi \, \frac{\phi^2}{m_{\rm P}^2}\right]\right)\,\phi^4 +\mathcal{O}(\phi^6).
\end{equation}
If the tree-level mass $m$ is much smaller than the Planck mass $m_{\rm P}$, then $\mathcal{O}(\phi^4)$ terms 
are strongly suppressed and do not influence the potential noticeably.
However, if $m\sim m_{\rm P}$, then $\mathcal{O}(\phi^4)$ terms can no longer be neglected, 
and the effective potential can develop new minima. However, this case hardly can be considered realistic, as the new minima are situated in the Planck region which lies beyond the model applicability domain. Therefore we only consider the case of the natural hierarchy $m \ll m_{\rm P}$.

Let us turn to a model with non-minimal interactions between gravity and the scalar field. 
According to the Horndeski Lagrangians~\eref{Horndeski_Lagrangians}, a model accounting 
for the simplest leading order interactions reads
\begin{equation}\label{the_non-minimal_action}
  \fl\mathcal{A}_1=\int d^4 x\sqrt{-g}\left[-\frac{2}{\kappa^2}  R+\frac12\,\left(g^{\mu\nu} +\beta\, G^{\mu\nu} \right)\, \pd_\mu\phi\, \pd_\nu\phi-\frac12\, \left( m^2+\lambda R\right)\,\phi^2 \right].
\end{equation}
Here $R$ is the scalar curvature and $G_{\mu\nu}$ is the Einstein tensor. Terms $R\,\phi^2$ and $G^{\mu\nu}\,\pd_\mu\phi\,\pd_\nu\phi$ describe new non-minimal three-particle
interactions. Term $G^{\mu\nu}\,\pd_\mu\phi\,\pd_\nu\phi$ is known as the John interaction from 
the Fab Four class~\cite{Charmousis:2011bf}, while $R\,\phi^2$ is typical for Brans-Dicke-like models 
with conformal symmetry~\cite{Chernikov:1968zm,Deser:1970hs}. The mass term also provides an additive 
contribution to the effective potential that was found above, so we will not consider it further.

In full analogy with the minimal coupling case, the John interaction does not contribute to an 
effective potential which makes {\em the third important result of the paper}. As the John term is 
bilinear in derivatives, the corresponding interaction vertex vanishes if at least one scalar field 
carries a zero momentum. This makes a complete analogy to the minimal coupling case.

The Brans-Dicke-like interaction $R\, \phi^2$, on the contrary, contributes to the effective potential. 
The only part of the interaction relevant in the given setup reads
\begin{equation}
    \int\, d^4 x \, \sqrt{-g}\left[ -\frac{\lambda}{2}\, R\,\phi^2 \right]\to\, \int d^4 x\left[-\lambda\, \kappa\, h^{\mu\nu}~\phi(\pd_\mu\pd_\nu - \eta_{\mu\nu}\square)\phi \, \right].
\end{equation}
As in the minimal coupling case, it is impossible to construct a vertex function with an odd number 
of fields, as the corresponding contributions vanish. An arbitrary vertex function with an even number 
of fields is given by
\begin{equation}
    \Gamma_{2n}=\left(\frac12\,\lambda^2 \,\kappa^2\,(3-d)(d-1)\right)^n~\mu^{4-d} \int\frac{d^dk}{(2\pi)^d}\,.
\end{equation}
In contrast to the minimal coupling case, such vertex functions have a much simpler momentum structure, 
so the corresponding integral can be evaluated explicitly and it is equal to the regularized volume of 
the momentum space. Consequently, the effective potential reads
\begin{eqnarray}\label{the_second_effective_potential}
  \eqalign{ (V_1)_{\rm eff} &=\frac{1}{2}\,\ln\left[1+ \frac{(d-1)(d-3)}{2} \, \lambda^2\, \kappa^2 \phi^2\right]~ \left( -i\, \mu^{4-d}\, \int\frac{d^dk}{(2\pi)^d} \right)  \cr
    &=\frac12\, \ln\left[ 1+\frac{3}{2}\,\lambda^2\,\kappa^2\,\phi^2 \right] \left(\int\frac{d^4_E k}{(2\pi)^4}\right).}
\end{eqnarray}
It should be noted that similar terms are usually omitted if evaluated in dimensional regularization. 
However, we believe that they should not be disregarded that easily, as they are relevant for the 
naturalness of a model~\cite{Jack:1990pz,Jack:1989tv}. 

The presence of a non-vanishing mass term is crucial for the effective potential in full analogy with 
the previous case. The complete effective potential generated by $\mathcal{A}_1$ consists of two parts:
\begin{equation}
  V_{\rm eff}=(V_0)_{\rm eff} + (V_1)_{\rm eff}.
\end{equation}
In the massless case $m=0$, the contribution $(V_0)_{\rm eff}$ vanishes. 
The potential~\eref{the_second_effective_potential}, on the contrary, develops a non-vanishing mass-like contribution that should be renormalized:
\begin{equation}
  (V_1)_{\rm eff}=\frac12\,\left(\frac32\,\lambda^2\,\kappa^2\,\int \frac{d^4_E k}{(2\pi)^4}\right) \,\phi^2 -\frac{9}{16} \,\kappa^4 \lambda^4 \phi^4 \,\int\frac{d^4_E k}{(2\pi)^4} + \mathcal{O}(\phi^6).
\end{equation}
There are two possible ways to perform the renormalization. The first one is to respect the structure 
of the tree-level theory and to normalize $(V_1)_{\rm eff}$ on the vanishing mass. In this case 
the whole potential vanishes in full analogy with the previous case. The second option is to assume 
a spontaneous dynamical generation of a new energy scale which produces the scalar field mass at 
the level of radiation corrections similarly to~\cite{Coleman:1973jx}. Therefore the mass term 
generated by $(V_1)_{\rm eff}$ should be normalized on a non-vanishing value of the mass 
$m_{\rm obs}$ measured in the low-energy regime $E\sim 0$. % some energy level $\mu$.
The corresponding counterterm is
\begin{equation}
  \delta\mathcal{L}=\frac{m_{\rm obs}^2}{2}\,\left[1-\frac32\,\frac{\kappa^2 
  \lambda^2}{m_{\rm obs}^2}\,\int\frac{d^4_E k}{(2\pi)^4}\right]\,\phi^2.
\end{equation}
So, the renormalized potential reads
\begin{equation}\label{the_second_effective_potential_ren}
    (V_1)_{\rm eff,ren} = \frac12\,\ln\left[1+\frac32\,\lambda^2\,\kappa^2\phi^2\right]\, \left(\frac{2m_{\rm obs}^2}{3 \kappa^2 \lambda^2}\right).
\end{equation}

We discuss in detail the role of these scenarios in the next Section. In the rest of this Section 
we focus on the case of a non-vanishing mass. The complete effective potential should be normalized 
in a cut-off regularization scheme since $(V_1)_{\rm ren}$ is ill-defined within dimensional 
regularization. The compete effective potential reads
\begin{eqnarray}\label{the_third_effective_potential}
  \fl\eqalign{  V_{\rm eff} &=\frac{1}{2} \, \int\frac{d^4_E k}{(2\pi)^4} \, \ln\left[ 1+\frac{\kappa^2 \phi^2}{\frac{k^2}{m^2} \left(\frac{k^2}{m^2}+1\right)}\right] +\frac12\ln\left[1+\frac32\,\lambda^2\,\kappa^2\,\phi^2\right] \int\frac{d^4_E k}{(2\pi)^4} \cr
    &=\frac{1}{32\pi^2}\left[ \,m^4 \,\int\limits_0^{\Lambda^2/m^2}\,dx\,x\ln\left[1+\frac{\kappa^2\phi^2}{x(x+1)}\right] + \frac12\,\ln\left[1+\frac32\,\lambda^2\,\kappa^2\,\phi^2\right]\Lambda^4\right] \cr
    &=\frac{m^4}{32\pi^2}\Bigg[\frac12\, \frac{\Lambda^4}{m^4}\, \left\{ \ln\left[1+\frac{m^4 \kappa^2\phi^2}{ \Lambda^2 \left( \Lambda^2 +m^2\right)}\right] + \ln\left[1+\frac32\,\lambda^2\,\kappa^2\,\phi^2\right]\right\} \cr
      &~~~+ \frac{\kappa^2\phi^2}{2}\ln\left[ 1+ \frac{\Lambda^2(\Lambda^2+m^2)}{m^4\kappa^2\phi^2} \right] \cr
      &~~~-\frac{\kappa^2\phi^2}{\sqrt{1-4\kappa^2\phi^2}}\ln\left[\frac{2\Lambda^2/m^2 +1-\sqrt{1-4\kappa^2\phi^2}}{2\Lambda^2/m^2+1+\sqrt{1-4\kappa^2\phi^2}}\, \frac{1+\sqrt{1-4\kappa^2\phi^2}}{1-\sqrt{1-4\kappa^2\phi^2}}\right] \cr
      &~~~+\frac{\kappa^2\phi^2}{\sqrt{1-4\kappa^2\phi^2}}\,\frac{1}{1+\sqrt{1-4\kappa^2\phi^2}}\, \ln\left[\frac{m^2}{\Lambda^2+m^2}\,\left(1+\frac{2\Lambda^2/m^2}{1-\sqrt{1-4\kappa^2\phi^2}} \right)\right] \cr
      &~~~+\frac{\kappa^2\phi^2}{\sqrt{1-4\kappa^2\phi^2}}\,\frac{1}{1-\sqrt{1-4\kappa^2\phi^2}}\, \ln\left[\frac{m^2}{\Lambda^2+m^2}\,\left(1+\frac{2\Lambda^2/m^2}{1+\sqrt{1-4\kappa^2\phi^2}} \right)\right]\Bigg], }
\end{eqnarray}
with $\Lambda$ being the cut-off scale. %The formal expression for potential \eref{the_third_effective_potential} contain $\log$-functions, but they do not introduce non-localities.

%
% NEW PART
The potential \eref{the_third_effective_potential} can be renormalized via standard methods. As it was highlighted above, it is a direct sum of two contribution $V_0$ \eref{the_first_effective_potential} and $V_1$ \eref{the_second_effective_potential} describing minimal and non-minimal interaction correspondingly. Therefore it requires two normalization conditions. For the sake of simplicity we normalize the potential on the observed mass $m_{\rm obs}$ and on the observed four-particle interaction coupling $g_{\rm obs}$. The normalization point $\phi_0$ should be set below the normalization scale. Therefore the normalization conditons read:
\begin{equation}\label{the_normalization_conditions}
  \left.\frac{d^2 V_{\rm eff}}{d\phi^2}\right|_{\phi_0} = m_{\rm obs}, ~~~ \left.\frac{d^4 V_{\rm eff}}{d\phi^4}\right|_{\phi_0} = g_{\rm obs} .
\end{equation}

Neither \eref{the_first_effective_potential} nor \eref{the_second_effective_potential} develops new minima in the effective theory applicability domain, so it appears that $\phi=0$ can be used as a suitable normalization point. However, this is not so. Potential \eref{the_third_effective_potential} is regular a proximity of $\phi\sim 0$ and vanishes in $\phi=0$ limit. Due to the presence of a non-vanishing tree-level mass $m$, its second derivative is also regular in a proximity of $\phi=0$. But the fourth derivative diverges in $\phi=0$ limit. This behavior is typical for effective models and it appears even in simple cases \cite{Coleman:1973jx}.

This feature of the theory, however, only shows that $\phi_0=0$ is not a suitable normalization point, so one should use some non-vanishing $\phi_0$ instead. In the rest respect the standard normalization procedure can be performed. It is possible to provide explicit expressions for the normalization conditions \eref{the_normalization_conditions}. However, such expressions are pretty massive and inconclusive, so we omit them for the sake of brevity.
%
% NEW PART END

Potential \eref{the_third_effective_potential} constitutes {\em the fourth result of this paper}. In full analogy to the previous case, 
this potential is regular in $\phi\sim 0$ and it develops no new minima in the relevant domain $4 \kappa^2 \phi^2 \ll 1$.
Behavior of this potential can be analyzed. First of all, higher-order terms in $\kappa\phi$ are irrelevant for an effective theory, as they can only modified the potential in the Planck region, which lies beyond the effective theory applicability domain. Secondly, as it was discussed, potentials \eref{the_first_effective_potential} and \eref{the_second_effective_potential} do not develop new minima in the low energy area. Finally, the fourth derivative of the potential diverges in $\phi=0$ limit, so the theory has peculiar infrared behavior. Therefore gravitational corrections described by the effective potential \eref{the_third_effective_potential} vanishes in $\phi=0$ and can only introduce small corrections to scalar field behavior in $\phi \sim 0$ region.

This Section can be summarized as follows. Firstly, it was found that the scalar field mass plays 
an important role in a generation of an effective potential. If a model admits only a minimal 
coupling, then the effective potential can be generated only if $m\not= 0$. If a model admits 
a non-minimal coupling, then there is an opportunity to generate the scalar field mass dynamically. 
Still, a scalar field of a non-vanishing mass generates effective 
potential~\eref{the_third_effective_potential}. In the next Section we discuss various implications 
of these results.

\section{Implementations}\label{section_implementations}

The role of a scalar field mass found in the previous Section agrees with previous findings~\cite{Coleman:1973jx}. Firstly, the mass provides a natural scale for regularization of 
the infrared sector of a theory. Secondly in the considered models, mass is the only dimensional parameter relevant for the problem.

The role of the Brans-Dicke-like interaction $R\,\phi^2$ can be understood in a similar way. 
This interaction is quadratic in derivatives which should be attributed to the gravitational 
sector rather than to the scalar one. In other words, the interaction describes a momentum-independent
coupling of a scalar field to a metric gradient. This makes it conceivable to explain how this
interaction can generate a non-vanishing effective potential in the massless case. The complete
energy-momentum tensor of a scalar field, which admits a momentum-dependent part, generates a 
non-vanishing curvature of the spacetime. The corresponding metric gradient couples to 
the momentum-independent part of the scalar field energy-momentum tensor and generates an 
interaction that does not depend on the scalar field momentum explicitly. This reasoning gives grounds 
to believe that the effective potential~\eref{the_second_effective_potential} can be indeed generated
in the massless case. 

It may appear that a new energy scale associated with the scalar field mass will be generated, 
but we would argue that this is not so. It is well-known that the following action similar to $\mathcal{A}_1$ admits the conformal symmetry~\cite{Chernikov:1968zm,Deser:1970hs}:
\begin{equation}
    \mathcal{A}_{\rm conformal} = \int d^4 x \sqrt{-g} \left[ \frac16\,R\,\phi^2 +\frac12\,g^{\mu\nu}\, \pd_\mu\phi\, \pd_\nu\phi \right] .
\end{equation}
The non-minimal interaction present in~\eref{the_non-minimal_action} is similar to the interaction 
in the conformal-invariant model. In contrast to the conformal model, the symmetry in $\mathcal{A}_1$ 
is explicitly violated by a dimensional parameter $\kappa$ related to the Planck mass. Therefore,
$\mathcal{A}_1$ admits at leas one dimensional parameter which, in term, defines the divergent 
mass term.

Based on this reasoning we argue the following. Firstly, in the case of a non-vanishing scalar field 
its mass plays the role of an energy scale regularizing the infrared sector. Secondly, in the massless 
case a scalar field with a specific non-minimal coupling to gravity can develop a non-vanishing 
effective potential. The non-minimal coupling resembles a model with conformal symmetry, so the 
corresponding effective potential resembles a potential that may dynamically break the symmetry. 
However the conformal symmetry of the considered model is explicitly broken at the tree-level, 
so the similarity can hardly point to a deeper relation to the symmetry breaking.
Finally, in both of these cases the effective potentials remain regular in the infrared sector. 
They do not develop new minima within the effective theory applicability domain 
and do not alter the vacuum energy content of the model. These facts show that at the one-loop 
level the fundamental physical properties of the model is hardly altered. 

It is important to point to possible implications of these effects in the context of inflation 
and of the Chameleon screening mentioned in the Introduction. Firstly, we would like to comment upon 
the Chameleon screening. It requires a certain form of a scalar field potential and 
a non-minimal coupling to matter of the following form~\cite{Khoury:2003rn,Khoury:2003aq,Brax:2008hh,Burrage:2017qrf}:
\begin{equation}
    \mathcal{L}_{\rm int}\sim \, g^{\mu\nu}\, T_{\mu\nu} \, \exp\left[\,\beta\, \kappa\,\phi\,\right].
\end{equation}
For the models addressed in this paper no such couplings were introduces, therefore there is 
no room for the screening.

The interaction required for the screening is typically found in $f(R)$-gravity models given in 
the Einstein frame~\cite{DeFelice:2010aj,Sotiriou:2008rp}. It can be argued that there is room for 
the Chameleon screening dynamically generated at the level of radiation corrections, but this problem
lies far beyond the scope of this paper and should be discussed elsewhere. We would only like to note
that such a scenario looks unrealistic. Firstly, potentials~\eref{the_first_effective_potential_ren}
and~\eref{the_second_effective_potential_ren} do not admit a form suitable for the screening. 
Secondly, dimensional considerations disfavor such an opportunity. A characteristic energy scale
generated by the Earth mean density $\rho_\oplus\simeq 5 \times 10^3 \, {\rm kg/m}^3$ is 
$\varepsilon \simeq 3 \,{\rm MeV}$. Effects associated with quantum gravity, including the considered
effective potential, can hardly be considered relevant at such low energies. Thus the influence 
of discussed effective potentials can safely be considered negligible in a terrestrial environment.

Secondly, possible implications for inflation should be noted. It is natural to expect 
an inflationary expansion to occur in the strong field regime $\kappa \phi \sim 1$ which puts 
it on the edge of an effective theory applicability. Nonetheless, formal expressions for the 
potentials are well-defined up to $\kappa\phi \simeq 1/2$, so the corresponding effects may be 
correctly estimated by the effective theory.

For a certain range of parameters both~\eref{the_first_effective_potential_ren} 
and~\eref{the_second_effective_potential_ren} admit small slow-roll parameters $\eta$ and 
$\epsilon$~\cite{Linde:1983gd,Senatore:2016aui,Gorbunov:2011zzc}. As we mentioned before, the first
effective potential~\eref{the_first_effective_potential} is sensitive to the mass-scale hierarchy 
and the effective model is applicable if and only if $m_{\rm obs}\sim m$ and $m \ll m_{\rm P}$. 
In that case $\epsilon$ and $\eta$ are small in a proximity of $\kappa\phi \sim 1$. 
The second effective potential~\eref{the_second_effective_potential_ren} is free from such  hierarchy dependence
and it also admits $\epsilon$, $\eta \sim 0$ about the Planck scale.

We would like to highlight ones more that the discussed models admit small slow-roll parameters 
at the edge of effective theory applicability domain. Nonetheless, we suppose that inflation within 
this models should be studied, as they may describe a natural appearance of an inflationary cosmology 
in the spirit of~\cite{Starobinsky:1980te}.

\section{Discussion and Conclusions}\label{section_conclusion}

The following results are found in this paper.
{\em The first result} clarifies the structure of an effective scalar field potential within the 
minimal model~\eref{the_minimal_model}. In the massless case a scalar field does not receive 
additional contributions to the effective potential at the one-loop level due to the structure of 
the energy-momentum tensor. 
{\em The second result} specifies the structure of the minimal-model effective potential with 
a scalar field of a non-vanishing mass. In that case the effective potential is given 
by~\eref{the_first_effective_potential} and it can be renormalized on the observed 
mass~\eref{the_first_effective_potential_ren}.
{\em The third result} shows that the John interaction does not contribute to the effective potential 
in a way similar to the mass term. The interaction itself provides a minimal beyond general relativity 
three-particle interaction. However, the interaction is bilinear in scalar field derivatives so 
it fails to generate a contribution to the effective potential in full analogy with the massless case.
{\em The fourth result} is the effective scalar field potential~\eref{the_third_effective_potential} 
obtained within a model admitting second-order field equations which accounts for all possible 
three-particle interactions~\eref{the_non-minimal_action}. This potential accounts both for the scalar 
field mass term and for the non-minimal interaction.

The following features of these potentials are important in the context of realistic models. 
Firstly, all potentials do not alter the vacuum energy content of a model. They do not develop 
new minima within the applicability domain and also vanish in $\phi \to 0$ limit. Therefore 
the ground state together with the vacuum energy remains unchanged. Secondly, the Chameleon screening,
mentioned in the Introduction, can hardly be realized in such models. The effective potentials have 
a form which is not suitable for the screening and the studied models lack a non-minimal interaction 
essential for the screening. Inflationary scenarios, on the contrary, may be successfully realized. 
Both parts of the effective potential generated by a mass term~\eref{the_first_effective_potential} 
and by the interaction term~\eref{the_second_effective_potential} have areas with small slow-roll 
parameters. An opportunity to describe an inflationary expansion within such a setup requires an 
additional investigation and will be discussed elsewhere.

Finally, possible implications of these results for the Standard Models should be highlighted. 
The Standard Model contains the Higgs field which should be affected by gravitational corrections 
in a way similar to the case presented. The part of a scalar field effective potential generated by 
a mass term~\eref{the_first_effective_potential} is sensitive to the hierarchy of the Higgs and 
Planck scales. Consequently, the contribution can be omitted in models with a realistic hierarchy 
$M_{\rm Higgs} \ll M_{\rm Pl}$. The contribution~\eref{the_second_effective_potential} associated 
with the non-minimal interaction $R\, \phi^2$, on the contrary, may not be easily neglected. 
It is defined by the value of the corresponding coupling $\lambda$~\eref{the_non-minimal_action}. 
For large $\lambda \sim 10^{17}$ the characteristic dimensionless parameter 
$\lambda\,\kappa\,\phi \sim 1$ at $\phi\sim M_{\rm Higgs}$ and the corresponding influence 
cannot be neglected. The value of $\lambda$, on the other hand, is not completely free, as it can 
be constrained by the empirical Solar system data on PPN parameters. A more detailed discussion of 
such a combined constraint on the model lies beyond the scope of this paper and will be discussed 
elsewhere alongside other formal developments of the model.

This discussion shows perspective directions for the further studies. Namely, an application of 
the effective potential technique to the Higgs sector of the Standard Model in the context of 
gravity should be analyzed. Results of this paper show a qualitative behavior of simple scalar field,
but they can hardly be considered sufficient in the particular case of the Standard Model. To proceed
with this goal it is required to evaluate the gravitational contribution to the Higgs effective
potential generated by possible non-minimal interactions with gravity. It allows one to constrain the
corresponding corrections with the empirical particle physics data. The non-minimal interaction with 
gravity, on the other hand, should be independently constrained by the Solar system data. Therefore 
combined constraints on the model can be found and a more comprehensive conclusion can be drawn about 
the role of non-minimal gravitational interaction within Higgs physics.

Finally, the opportunity to describe the inflation a la~\cite{Starobinsky:1980te} within such an 
approach must also not be forgotten. As it was highlighted above, effective potentials admit areas 
with small slow-roll parameters, so possible inflationary scenarios should be studied. Some formal
developments also should be discussed. We used the direct perturbative calculations to evaluate the 
effective potential which put some constraints on the studied spectrum of models. More sophisticated 
techniques similar to those discussed in~\cite{Buchbinder:1992rb} and those based on renormalization group approach \cite{Elizalde:1993ee,Elizalde:1994im,Elizalde:1994ds,Elizalde:1993ew,Elizalde:1993qh,Elizalde:1994gv} may provide a simple way to analyze 
models with more intricate structures. 

\ack

The work (B.L.) was supported by the Foundation for the Advancement of Theoretical Physics and Mathematics “BASIS”.

\section*{References}
\bibliographystyle{iopart-num}
\bibliography{EPoSTG.bib}

\providecommand{\newblock}{}
\begin{thebibliography}{10}
\expandafter\ifx\csname url\endcsname\relax
  \def\url#1{{\tt #1}}\fi
\expandafter\ifx\csname urlprefix\endcsname\relax\def\urlprefix{URL }\fi
\providecommand{\eprint}[2][]{\url{#2}}

\bibitem{Horndeski:1974wa}
Horndeski G~W 1974 {\em Int. J. Theor. Phys.\/} {\bf 10} 363--384

\bibitem{Kobayashi:2011nu}
Kobayashi T, Yamaguchi M and Yokoyama J 2011 {\em Prog. Theor. Phys.\/} {\bf
  126} 511--529 (\textit{Preprint} \eprint{1105.5723})

\bibitem{Ishak:2018his}
Ishak M 2019 {\em Living Rev. Rel.\/} {\bf 22} 1 (\textit{Preprint}
  \eprint{1806.10122})

\bibitem{Berti:2015itd}
Berti E {\em et~al.\/} 2015 {\em Class. Quant. Grav.\/} {\bf 32} 243001
  (\textit{Preprint} \eprint{1501.07274})

\bibitem{Clifton:2011jh}
Clifton T, Ferreira P~G, Padilla A and Skordis C 2012 {\em Phys. Rept.\/} {\bf
  513} 1--189 (\textit{Preprint} \eprint{1106.2476})

\bibitem{Tretyakova:2017fbu}
Tretyakova D~A and Latosh B~N 2018 {\em Universe\/} {\bf 4} 26
  (\textit{Preprint} \eprint{1711.08285})

\bibitem{Linde:1983gd}
Linde A~D 1983 {\em Phys. Lett.\/} {\bf 129B} 177--181

\bibitem{Senatore:2016aui}
Senatore L 2017 {Lectures on Inflation} {\em {Theoretical Advanced Study
  Institute in Elementary Particle Physics}: {New Frontiers in Fields and
  Strings}\/} pp 447--543 (\textit{Preprint} \eprint{1609.00716})

\bibitem{Gorbunov:2011zzc}
Gorbunov D~S and Rubakov V~A 2011 {\em {Introduction to the theory of the early
  universe: Cosmological perturbations and inflationary theory}\/}

\bibitem{Khoury:2003rn}
Khoury J and Weltman A 2004 {\em Phys. Rev.\/} {\bf D69} 044026
  (\textit{Preprint} \eprint{astro-ph/0309411})

\bibitem{Khoury:2003aq}
Khoury J and Weltman A 2004 {\em Phys. Rev. Lett.\/} {\bf 93} 171104
  (\textit{Preprint} \eprint{astro-ph/0309300})

\bibitem{Brax:2008hh}
Brax P, van~de Bruck C, Davis A~C and Shaw D~J 2008 {\em Phys. Rev.\/} {\bf
  D78} 104021 (\textit{Preprint} \eprint{0806.3415})

\bibitem{Burrage:2017qrf}
Burrage C and Sakstein J 2018 {\em Living Rev. Rel.\/} {\bf 21} 1
  (\textit{Preprint} \eprint{1709.09071})

\bibitem{Coleman:1973jx}
Coleman S~R and Weinberg E~J 1973 {\em Phys. Rev.\/} {\bf D7} 1888--1910

\bibitem{Buchbinder:1992rb}
Buchbinder I, Odintsov S and Shapiro I 1992 {\em {Effective action in quantum
  gravity}\/}

\bibitem{Burgess:2003jk}
Burgess C~P 2004 {\em Living Rev. Rel.\/} {\bf 7} 5--56 (\textit{Preprint}
  \eprint{gr-qc/0311082})

\bibitem{Donoghue:1994dn}
Donoghue J~F 1994 {\em Phys. Rev.\/} {\bf D50} 3874--3888 (\textit{Preprint}
  \eprint{gr-qc/9405057})

\bibitem{Bjerrum-Bohr:2014zsa}
Bjerrum-Bohr N~E~J, Donoghue J~F, Holstein B~R, Planté L and Vanhove P 2015
  {\em Phys. Rev. Lett.\/} {\bf 114} 061301 (\textit{Preprint}
  \eprint{1410.7590})

\bibitem{Heisenberg:2020cyi}
Heisenberg L, Noller J and Zosso J 2020  (\textit{Preprint}
  \eprint{2004.11655})

\bibitem{Calmet:2018qwg}
Calmet X and Latosh B 2018 {\em Eur. Phys. J.\/} {\bf C78} 205
  (\textit{Preprint} \eprint{1801.04698})

\bibitem{Latosh:2018xai}
Latosh B 2018 {\em Eur. Phys. J.\/} {\bf C78} 991 (\textit{Preprint}
  \eprint{1812.01881})

\bibitem{Arbuzov:2017nhg}
Arbuzov A~B and Latosh B~N 2017 {\em Eur. Phys. J.\/} {\bf C77} 702
  (\textit{Preprint} \eprint{1703.06626})

\bibitem{Zumalacarregui:2013pma}
Zumalacárregui M and García-Bellido J 2014 {\em Phys. Rev. D\/} {\bf 89}
  064046 (\textit{Preprint} \eprint{1308.4685})

\bibitem{Bekenstein:1992pj}
Bekenstein J~D 1993 {\em Phys. Rev. D\/} {\bf 48} 3641--3647 (\textit{Preprint}
  \eprint{gr-qc/9211017})

\bibitem{Kobayashi:2019hrl}
Kobayashi T 2019 {\em Rept. Prog. Phys.\/} {\bf 82} 086901 (\textit{Preprint}
  \eprint{1901.07183})

\bibitem{Barvinsky:1985an}
Barvinsky A~O and Vilkovisky G~A 1985 {\em Phys. Rept.\/} {\bf 119} 1--74

\bibitem{Rivat:2019xrq}
Rivat S 2019 {\em Stud. Hist. Phil. Sci. B\/} {\bf 68} 23--39

\bibitem{Barvinsky:1993zg}
Barvinsky A, Kamenshchik A and Karmazin I 1993 {\em Phys. Rev. D\/} {\bf 48}
  3677--3694 (\textit{Preprint} \eprint{gr-qc/9302007})

\bibitem{Shapiro:1995yc}
Shapiro I~L and Takata H 1995 {\em Phys. Rev. D\/} {\bf 52} 2162--2175
  (\textit{Preprint} \eprint{hep-th/9502111})

\bibitem{Steinwachs:2011zs}
Steinwachs C~F and Kamenshchik A~Y 2011 {\em Phys. Rev. D\/} {\bf 84} 024026
  (\textit{Preprint} \eprint{1101.5047})

\bibitem{Kamenshchik:2014waa}
Kamenshchik A~Y and Steinwachs C~F 2015 {\em Phys. Rev. D\/} {\bf 91} 084033
  (\textit{Preprint} \eprint{1408.5769})

\bibitem{Latosh:2020zho}
Latosh B 2020 {\em Phys. Part. Nucl.\/} {\bf 51} 859--878

\bibitem{Charmousis:2011bf}
Charmousis C, Copeland E~J, Padilla A and Saffin P~M 2012 {\em Phys. Rev.
  Lett.\/} {\bf 108} 051101 (\textit{Preprint} \eprint{1106.2000})

\bibitem{Chernikov:1968zm}
Chernikov N and Tagirov E 1968 {\em Ann. Inst. H. Poincare Phys. Theor. A\/}
  {\bf 9} 109

\bibitem{Deser:1970hs}
Deser S 1970 {\em Annals Phys.\/} {\bf 59} 248--253

\bibitem{Jack:1990pz}
Jack I and Jones D 1990 {\em Nucl.Phys.B\/} {\bf 342} 127--148

\bibitem{Jack:1989tv}
Jack I and Jones D 1990 {\em Phys. Lett. B\/} {\bf 234} 321--323

\bibitem{DeFelice:2010aj}
De~Felice A and Tsujikawa S 2010 {\em Living Rev. Rel.\/} {\bf 13} 3
  (\textit{Preprint} \eprint{1002.4928})

\bibitem{Sotiriou:2008rp}
Sotiriou T~P and Faraoni V 2010 {\em Rev. Mod. Phys.\/} {\bf 82} 451--497
  (\textit{Preprint} \eprint{0805.1726})

\bibitem{Starobinsky:1980te}
Starobinsky A~A 1980 {\em Phys. Lett.\/} {\bf B91} 99--102

\bibitem{Elizalde:1993ee}
Elizalde E and Odintsov S 1994 {\em Russ. Phys. J.\/} {\bf 37} 25--29
  (\textit{Preprint} \eprint{hep-th/9302074})

\bibitem{Elizalde:1994im}
Elizalde E and Odintsov S 1994 {\em Phys. Lett. B\/} {\bf 333} 331--336
  (\textit{Preprint} \eprint{hep-th/9403132})

\bibitem{Elizalde:1994ds}
Elizalde E, Kirsten K and Odintsov S 1994 {\em Phys. Rev. D\/} {\bf 50}
  5137--5147 (\textit{Preprint} \eprint{hep-th/9404084})

\bibitem{Elizalde:1993ew}
Elizalde E and Odintsov S 1994 {\em Phys. Lett. B\/} {\bf 321} 199--204
  (\textit{Preprint} \eprint{hep-th/9311087})

\bibitem{Elizalde:1993qh}
Elizalde E and Odintsov S 1994 {\em Z. Phys. C\/} {\bf 64} 699--708
  (\textit{Preprint} \eprint{hep-th/9401057})

\bibitem{Elizalde:1994gv}
Elizalde E, Odintsov S~D and Romeo A 1995 {\em Phys. Rev. D\/} {\bf 51}
  1680--1691 (\textit{Preprint} \eprint{hep-th/9410113})

\end{thebibliography}

\end{document}